\documentstyle[prc,epsfig,aps]{revtex}
\begin{document}

\title{
POLARIZATION DEGREES OF FREEDOM  IN TWO-NUCLEON KNOCKOUT 
 FROM FINITE NUCLEI}

\author{Jan Ryckebusch 
\footnote{Talk presented by Jan Ryckebusch at
the Fourth Workshop on Electromagnetically Induced Two-Hadron
Emission, Granada (Spain), May 26-29, 1999.} 
and Wim Van Nespen}
\address{
Department of Subatomic and Radiation Physics\\
University of Gent \\ Proeftuinstraat 86, B-9000 Gent, Belgium}

\maketitle

\begin{abstract}
Polarization observables for the $A(e,e'NN)$ and $A(\gamma,NN)$
reactions are a powerful tool to study nucleon-nucleon correlations in
the nuclear medium.  In this paper, model calculations for the
$^{4}$He($e,e'pp)$, $^{16}$O($\vec{e},e' \vec{p}p)$,
$^{16}$O($\vec{e},e' \vec{p}n)$ and $^{12}$C($\vec{\gamma},\vec{p}N$)
reactions are presented.  The sensitivity of the differential cross
sections and polarization observables to central and spin-dependent
nucleon-nucleon correlations is discussed.
\end{abstract}

\section{Introduction}

The holy grail in the study of electromagnetically induced two-nucleon
knockout processes is the detection of signals and study of the
short-range correlations in the nuclear medium.  Eventhough the latter
are generally accepted to be indispensable for the binding of nuclei,
a direct experimental confirmation of their existence is still
lacking.  Our goal here is to show that polarization degrees of
freedom in $A(e,e'NN)$ and $A(\gamma,NN)$ studies offer good
opportunities to probe and study nucleon-nucleon correlations. To do
this we present results of model calculations in a non-relativistic
framework.  We start from a shell-model formulation for the two-nucleon
knockout reaction process, implement corrections for central and
spin-dependent correlations and then show how important these
correlations are for $A(\gamma,NN$) and $A(e,e'NN)$ differential cross
sections and polarization observables.  These studies are conducted in
a model which also accounts for the ``background'' of meson-exchange
and $\Delta$-isobar currents.  These conventional two-nucleon currents
are unmistakingly a strong source of two-nucleon knockout strength and
complicate the interpretation of the data in terms of nucleon-nucleon
correlations.

The paper is organized as follows.  In Section 2 we briefly review the
two-nucleon knockout observables in terms of structure functions and
discuss their dependence on the different observables.  In Section 3
we present a framework that aims at extracting from the two-nucleon
knockout observables the information on the
nucleon-nucleon correlations.  Section 4 discusses the model which was
employed to calculate the cross sections and polarization observables.
In Section 5 we apply this machinery to the calculation of
$^{16}$O($\vec{e},e' \vec{p}p)$, $^{16}$O($\vec{e},e' \vec{p}n)$ and
$^{12}$C($\vec{\gamma},\vec{N}N$) observables.

\section{Structure functions and polarization observables in
  ($e,e'NN$) and ($\gamma,NN$)}
The ($e,e'NN$) differential cross section for excitation of specific
states in the residual nucleus can be
cast in the form
\begin{eqnarray*}
& & {d^8 \sigma  \over dE_1 d \Omega _1 d \Omega _2 d \epsilon ' d \Omega
_{\epsilon '}} (\vec{e},e'N_1N_2)  =  
{1 \over 4 (2\pi)^8 } p_1 p_2 E_1 E_2 f_{rec}^{-1} \sigma_{M}
\nonumber \\ 
\times
& & \Biggl[ v_T {W_T}
+ v_L {W_L}
+ v_{LT} {W_{LT}}
+ v_{TT} {W_{TT}} 
+ h \biggl[ v'_{LT} {W'_{LT}} + 
v'_{TT} {W'_{TT}} \biggr] \Biggr] \; .
\label{eq:eepnn}
\end{eqnarray*}

\begin{figure}
\begin{center}
\includegraphics[totalheight=5.cm,width=12.5cm]{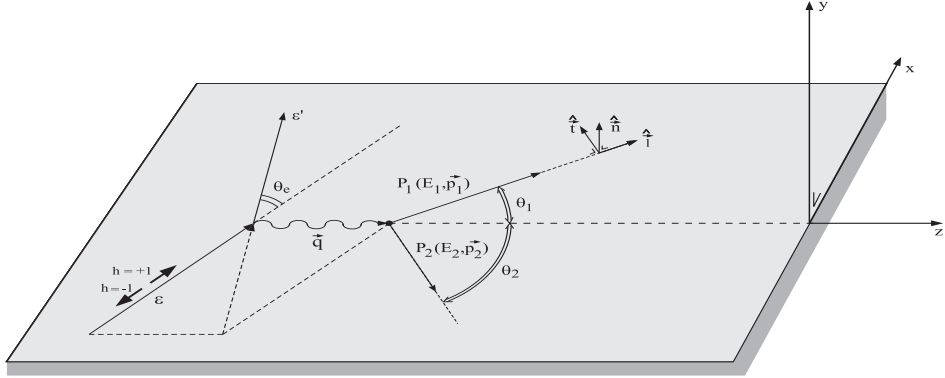}
\end{center} 
\caption{\sl Kinematic definitions for the $A(\vec{e},e' \vec{N}N$)
reaction in coplanar kinematics.}
\label{fig:conventions}
\end{figure}

Here, all structure functions {$W's$} depend on 
the variables {(q,$\omega$,p$_1$,p$_2$,$\theta _1$,$\theta_2$ and $\phi _1 - \phi
_2$)} in an non-trivial manner.  In addition, the structure functions { $W_{LT}, W_{TT}$ and
$W'_{LT}$}  depend  on the azimuthal angle of the
center-of-mass (c.o.m.) of the active nucleon pair $\Phi \equiv \frac{\phi _1 + \phi_2}
{2}$.  This dependence reflects itself in factors of the form $\sin(2\Phi)$ and
$\cos(2 \Phi)$ that can be formally pulled out of the structure functions.
The question arises whether the availability of polarized electrons is
a great asset in two-nucleon knockout  studies.  For symmetry reasons, the structure
function {$W'_{LT}$} is identical zero in coplanar
kinematics and {$W'_{TT}$}=0.  Additional observables
come into reach of investigation when considering the possibility of
performing recoil polarization measurements on
one of the ejected nucleons.  We express the recoil polarization in
the reference frame determined by the unit vectors (Figure \ref{fig:conventions}) 
\begin{displaymath}
\hat{\vec{l}} = \frac {\vec{p}_1} {\left| \vec{p}_1 \right|}
\; \; \; \; \;
\hat{\vec{n}} = \frac {\vec{q} \times \vec{p}_1} 
{\left| \vec{q} \times \vec{p}_1 \right|}
\; \; \; \; \;
\hat{\vec{t}} = \hat{\vec{n}} \times \hat{\vec{l}} \; .
\end{displaymath} 

Recoil polarization observables can be determined through measuring {\bf ratios}
\begin{eqnarray*}    
{\mathrm P}_i &=&  \frac {\sigma (s_{1i}\uparrow) -  
\sigma (s_{1i}\downarrow)} 
{\sigma (s_{1i}\uparrow) + \sigma (s_{1i}\downarrow)} \nonumber \\
 {\mathrm P}_i' &=& \frac {
\left[ \sigma (h=1,s_{1i}\uparrow) - \sigma (h=-1,s_{1i}\uparrow) \right] 
- \left[ \sigma (h=1,s_{1i}\downarrow) - 
\sigma (h=-1,s_{1i}\downarrow) \right] }
{
\left[ \sigma (h=1,s_{1i}\uparrow) + \sigma (h=-1,s_{1i}\uparrow) \right] 
+ \left[ \sigma (h=1,s_{1i}\downarrow) +  
\sigma (h=-1,s_{1i}\downarrow) \right] }
\end{eqnarray*}
where {$s_{1i}\uparrow$} denotes that nucleon ``1'' is spin-polarized
in the positive i direction i=(n,l,t). For the sake of completeness we
mention that for real photons the differential cross section and
photon asymmetry can be written as
\begin{eqnarray*}
\frac {d^4 \sigma}
{d \Omega _1 d \Omega _2 dE_1 dE_2}
&=& \frac {1} {(2\pi)^5 2 E_{\gamma}} p_1 p_2 E_1 E_2 
\delta (E_{A-2} + E_1 + E_2 - E_A - E_{\gamma}) \frac {1} {2}
{W_T} 
\\
\Sigma &=& 
 \frac {d \sigma _{\parallel} (\vec{\gamma},NN) -
                d \sigma _{\perp} (\vec{\gamma},NN) }
                {d \sigma _{\parallel} (\vec{\gamma},NN) +
                d \sigma _{\perp} (\vec{\gamma},NN) } = 
-  {\frac {W_{TT}} {W_{T}}} \; ,
\label{eq:sigma}    
\end{eqnarray*}

\section{Central and spin-dependent correlations in
  ($e,e'NN$) and ($\gamma,NN$)}

Two-nucleon knockout observables have the potential of providing
information on the following transition matrix elements
\begin{equation}
m_F^{fi} (\lambda = \pm 1,0)
 =  
< \widetilde{\Psi_f}(E_f)\mid  {J_{\pm 1,0}^{[1]}+J_{\pm 1}^{[2]}} \mid
 \widetilde{\Psi_i}(E_i)> \; ,
\label{eq:matrixelement}
\end{equation}
where the exclusive character of the reaction imposes the final state
to be of the form
\begin{equation}
\left| \Psi _f \right> = 
\left| \Psi_{f}^{A-2}(E_x,J_RM_R);
{\vec p}_1 m_{s_{1}}{\vec p}_2 m_{s_{2}} \right> \; ,
\end{equation}
and $\left| \Psi_i \right>$ is the ground-state of the target
nucleus.  A convenient and widely used way of writing correlated wave functions is
\begin{displaymath}
\mid \widetilde{\Psi} > = {\cal S} \left[\prod _{i<j=1} ^{A} \left(
1 - {g_c(r_{ij})+f_{t\tau}(r_{ij})\widehat{S_{ij}} \vec{\tau}_i
. \vec{\tau}_j 
+ f_{\sigma \tau}(r_{ij}) \vec{\sigma}_i \vec{\sigma}_j  \vec{\tau}_i
. \vec{\tau}_j }
\right) \right] \mid \Psi > \nonumber \; ,
\end{displaymath}
where ${\cal S}$ is a symmetrization operator and $\mid \Psi >$ is the
Slater determinant wave function as it can be determined in an
independent particle model (IPM) for the nucleus.  The functions
$g_c$, $f_{t\tau}$ and $f_{\sigma \tau}$ are the central, tensor and
spin-isospin correlation functions.  They are expected to be nearly
mass independent and therefore represent an universal feature
of nuclei \cite{benhar93}.  
In our calculations, the one-body current operator $J_{\pm 1,0}^{[1]}$
in Eq.~\ref{eq:matrixelement} is considered in the non-relativistic
impulse approximation.  For the two-body contributions $J_{\pm 1}^{[2]}$
we consider the conventional pion-exchange (MEC) and include also
$\Delta$-isobar current terms.

We restrict ourselves to exclusive two-nucleon knockout reactions.
The cross sections are computed in the so-called spectator
approximation which in some sense is equivalent with a lowest-order
cluster expansion retaining all operators up to the two-body
level. For inclusive reactions this approach was shown to be a 
reasonable approximation (see e.g. Refs. \cite{co98} and
\cite{leidemann94}).  For exclusive reactions, it is expected to be an
even better approximation as the exclusive character of the process is
heavily selective with respect to the number of nucleons that get
involved in the reaction process and three- and more-body operators
are likely to play only a marginal role.  In the spectator
approximation, one ends up with the following two contributions to the
transition matrix elements.  First, there is the contribution from the
two-body currents that produces two-nucleon strength even in the
strict IPM limit

\vspace{0.5cm}
{\large \sl
UNCORRELATED PART}
\begin{displaymath}
<\Psi_f (E_f) \mid {J^{[2]}_{\mu}(q)} 
\mid {\Psi_i}(E_i)> \nonumber \\
\end{displaymath}

Second, the information wirth regard to the nucleon-nucleon
correlations comes from  

\vspace{0.5cm}
{\large \sl CORRELATED PART}

\begin{displaymath}
<\Psi_f  \mid   \sum _{i<j=1} ^{A} \Biggl[ 
\Biggl( {J^{[1]}_{\mu}(i;q) + J^{[1]}_{\mu}(j;q) + 
J^{[2]}_{\mu}(i,j;q)} \Biggr) 
 {\Biggl( - g_c(r_{ij})+f_{t\tau}(r_{ij})\widehat{S_{ij}} \vec{\tau}_i
. \vec{\tau}_j  + 
f_{\sigma \tau}(r_{ij}) \vec{\sigma}_i \vec{\sigma}_j  \vec{\tau}_i
. \vec{\tau}_j \Biggr)}
+ h.c. \Biggr]
\mid {\Psi_i}> 
\end{displaymath}

\begin{figure}
\begin{center}
\includegraphics[bb=10 380 582 753,totalheight=8.cm,width=10.cm]{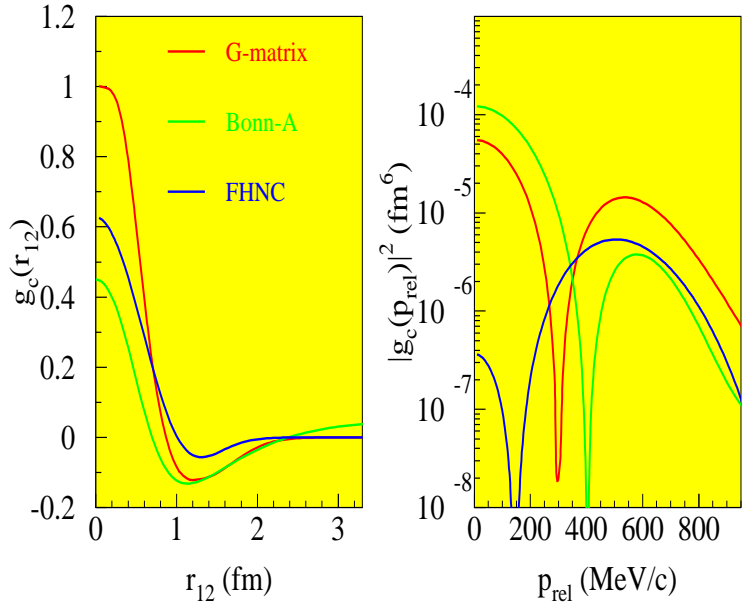}
\end{center}
\caption{\sl The central correlation function in coordinate and momentum
space as calculated in various many-body theories.} 
\label{fig:centralf}
\end{figure}

A transparent way of connecting the $A(e,e'NN$) observables to the
information on the nucleon-nucleon correlations is provided by the
{\sl FACTORIZED APPROACH } as it was e.g. formulated in
Ref.~\cite{ryckebusch96}.  In this semi-analytical scheme, that
admittely involves rather severe approximations, a strict separation
between the relative and c.o.m. momentum of the active pair in the
expression for the $A(e,e'pp)$ differential cross sections can be
attained

\begin{equation}
\frac {d^5 \sigma} {d \epsilon ' d \Omega _{\epsilon '} d \Omega _{1} 
 d \Omega _{2} dE _{1}}
       = E_1 p_1 E_2 p_2 f_{rec} ^{-1} \sigma _{epp} \left( 
              k _{+} , k  _{-} ,q        \right)
F_{h_{1},h_{2}}(P) \; ,
\label{eq:factorized}
\end{equation}

where the relative $(k_{\pm})$ and c.o.m. momentum $(P)$ of the active pair in the
initial state are determined from the measured momenta by  
\begin{eqnarray}
\vec{p}_{rel}  & =  & \vec{k}_{\pm} = \frac {\vec{p}_{1} -\vec{p}_{2}} {2}  
\pm \frac {\vec{q}} {2}  \nonumber \\ 
\vec{P} & = & \vec{p}_{1} + \vec{p}_{2} - \vec{q}
\end{eqnarray} 
and the information on the relative motion of the ejected proton
pair is contained in
\begin{equation}
\sigma _{epp}  =  \sigma _ M  
\biggl[ \frac {q _{\mu} ^4} {q ^4} {w_L} + 
 \left( 
          \frac {-q _{\mu} ^2} {2q ^2} + tan ^2 \frac {\theta _e} {2}
\right)
w_T 
 + \frac {q _{\mu} ^2} {2q ^2} w_{TT} + 
\frac {1} {\sqrt{2}} \frac {q _{\mu} ^2} {q ^3} (\epsilon + \epsilon
') tan \frac {\theta _e} {2} w_{LT} \biggr] \; .
\end{equation}
\noindent
For example, in the spectator approximation the longitudinal
contribution $w_L$ can be cast in the form
\begin{eqnarray*}
{w_L} & = & 4e^2  \left( g_c(k_+) + g_c(k_-) \right)^2
\left( G_E(q_{\mu}q^{\mu}) \right) ^2
\nonumber \\ & & + 40 e^2  \left( 
f_{\sigma \tau}(k_+) + f_{\sigma \tau}(k_-) \right)^2
\left( G_E(q_{\mu}q^{\mu}) \right) ^2 
\nonumber \\
& &
+ 24 e^2  \left( g(k_+) + g(k_-) \right) 
\left( f_{\sigma \tau} (k_+) + f_{\sigma \tau} (k_-) \right)
\left( G_E(q_{\mu}q^{\mu}) \right) ^2 
\nonumber \\
& & 
+ \frac {16} {3} \sqrt{ \frac {\pi} {5}} e^2 
\left( g(k_+) + g(k_-) \right) 
\left( f_{t \tau}^0 (- \vec{k} _+) + f_{t \tau}^0 (-\vec{k}_-) \right)
\left( G_E(q_{\mu}q^{\mu}) \right) ^2 \; .
\end{eqnarray*}
\noindent
This expression nicely illustrates that the
correlation functions $g_c$, $f_{t\tau}$ and  $f_{\sigma \tau}$
establish the link between two-nucleon knockout observables and the
physics beyond mean-field behaviour of nuclei.

As illustrated in Figure \ref{fig:centralf} the various many-body
theories produce vastly different predictions for the central
correlation function $g_c$.  The G-matrix calculation in nuclear
matter with the Reid potential by W.H. Dickhoff and C. Gearhart
\cite{gearhart} is the only calculation contained in Figure
\ref{fig:centralf} that produced a correlation function with a hard
core at short distances. This correlation function produced a
favorable agreement with the $^{12}$C($e,e'pp$) data from MAMI \cite{blom98}. 

\begin{figure}
\begin{center}
\includegraphics[bb=10 380 582 780,totalheight=6.cm,width=10.cm]{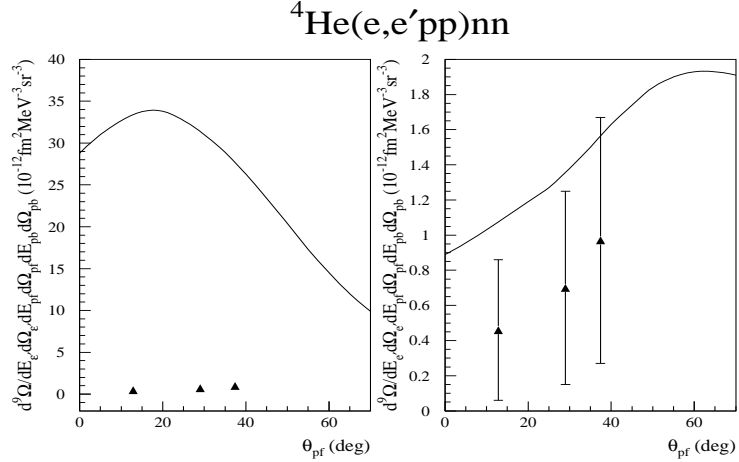}
\end{center}
\caption{\sl Dependence of the $^{4}$He$(e,e'pp)$ differential cross
section on the polar angle of the forward emitted proton $(\equiv
\theta_{pf})$. The solid curve in the left (right) panel is obtained
by using the OMY (G-matrix) correlation function in the factorized
model of Ref.~\protect \cite{ryckebusch96}.  With increasing $\theta_{pf}$ the
relative momentum of the ejected diproton grows. The data are from
Ref.~\protect \cite{devries}.}
\label{fig:helium}
\end{figure}

Figure \ref{fig:helium} illustrates for the $^4$He$(e,e'pp)$ case the
sensitivity of the measured differential cross sections to the choice
of the correlation function. The curves were obtained with the
factorized formulae of Eq.~(\ref{eq:factorized}) and represent a
weighted average over the experimental phase space.  The $^4$He data
from Ref. \cite{devries} seem to prefer the G-matrix correlation
function and clearly exclude a hard-core correlation function like the
OMY one.
Other correlation functions included in Figure \ref{fig:centralf} are
the ones obtained by FHNC calculations for $^{16}$O by F.Arias de
Saavedra et al. \cite{arias} and the dressed RPA calculation by
W. Geurts {\em et al.} \cite{geurts} with the Bonn-A potential.  The
latter correlation function is extremely soft at short distances and
predicts strong central correlations at intermediate distances (1-2
fm).  In momentum space this behaviour reflects itself in little high
relative momenta components.  It is foreseen that continued
two-nucleon knockout investigations will eventually be able to
fully map the correlation functions, thereby establishing
the origin of the high-momentum components in the nuclear wave
functions.

\begin{figure}
\begin{center}
\includegraphics[bb=10 11 580 390,totalheight=8.cm,width=10.cm]{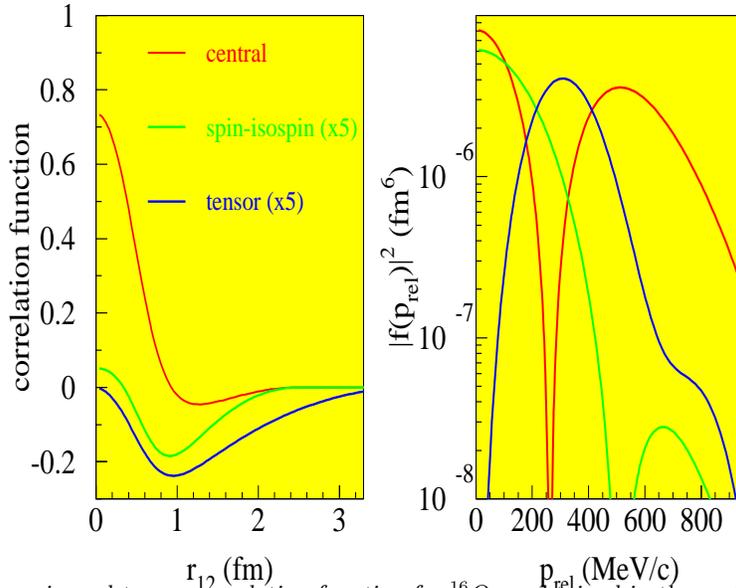}
\caption{\sl The central, spin-isospin and tensor correlation function for
$^{16}$O as obtained in the variational calculation of Ref.~\protect \cite{pieper}.}
\end{center}
\label{fig:pieper}
\end{figure}

Figure 4 shows the central and spin-dependent
correlation functions as they were obtained in the variational
calculations for the ground-state of $^{16}$O with the Argonne
$v_{14}$ NN potential \cite{pieper}.  In coordinate space the spin-isospin and
tensor correlation function appear considerably weaker than the
central one (note that they were multiplied with a factor of five in
the left panel of Figure 4).  In momentum space a very
different picture emerges.  For relative momenta in the region 200-400
MeV/c the spin-dependent correlation functions overshoot the central
correlation function.   For $p_{rel} >$ 500~MeV/c  the
central correlations  dominate.  Using the factorized approach of
Eq. (\ref{eq:factorized}) in combination with the central and
state-dependent correlation functions of Fig.~4 no
favorable agreement with the $^{12}$C($e,e'pp$) data from
Ref.\cite{blom98} could be reached.  This might be  due to the fact
that the central correlation function obtained in the variational
calculations is too soft.

\section{Theory}
In order to extract from the two-nucleon knockout data information on
the nucleon-nucleon correlations, a framework is required in which
besides the NN correlations also the final-state interaction and the
conventional (meson-exchange and $\Delta$-isobar) two-body currents
are accounted for.  We have developed such a formalism \cite{jan97}.
It is based on a consistent shell-model description for the inital and
final state.  This approach has the advantage of respecting the
orthogonality and antisymmetry of all the nuclear wave functions
involved in the calculation.  A distorted wave description of the two
ejected nucleons is adopted and the differential cross section and
polarization observables can be calculated for excitation of each
discrete state in the spectrum of the final nucleus.  In constructing
the nuclear Slater determinants a realistic set of single-particle
wave functions is used.  This makes a formal separation of the
relative and c.o.m. motion impractical as this would require an
additional expansion of the wave functions in terms of a harmonic
oscillators.  One of greatest difficulties in performing exclusive
two-nucleon knockout calculations is the dimensionality of the
problem.  In contrast to calculations of cross sections for inclusive
processes like $A(e,e')$, no closure relations can be used to reduce
the dimensionality of the problem. We have exploited partial wave
expansion and Racah algebra techniques to analytically reduce the
dimension of the matrix elements which occur when considering three
particles (two ejectiles and the residual nucleus) in the final state.

\section{Results and discussion}

\begin{figure}
\begin{center}
\includegraphics[totalheight=12.5cm,width=10.cm]{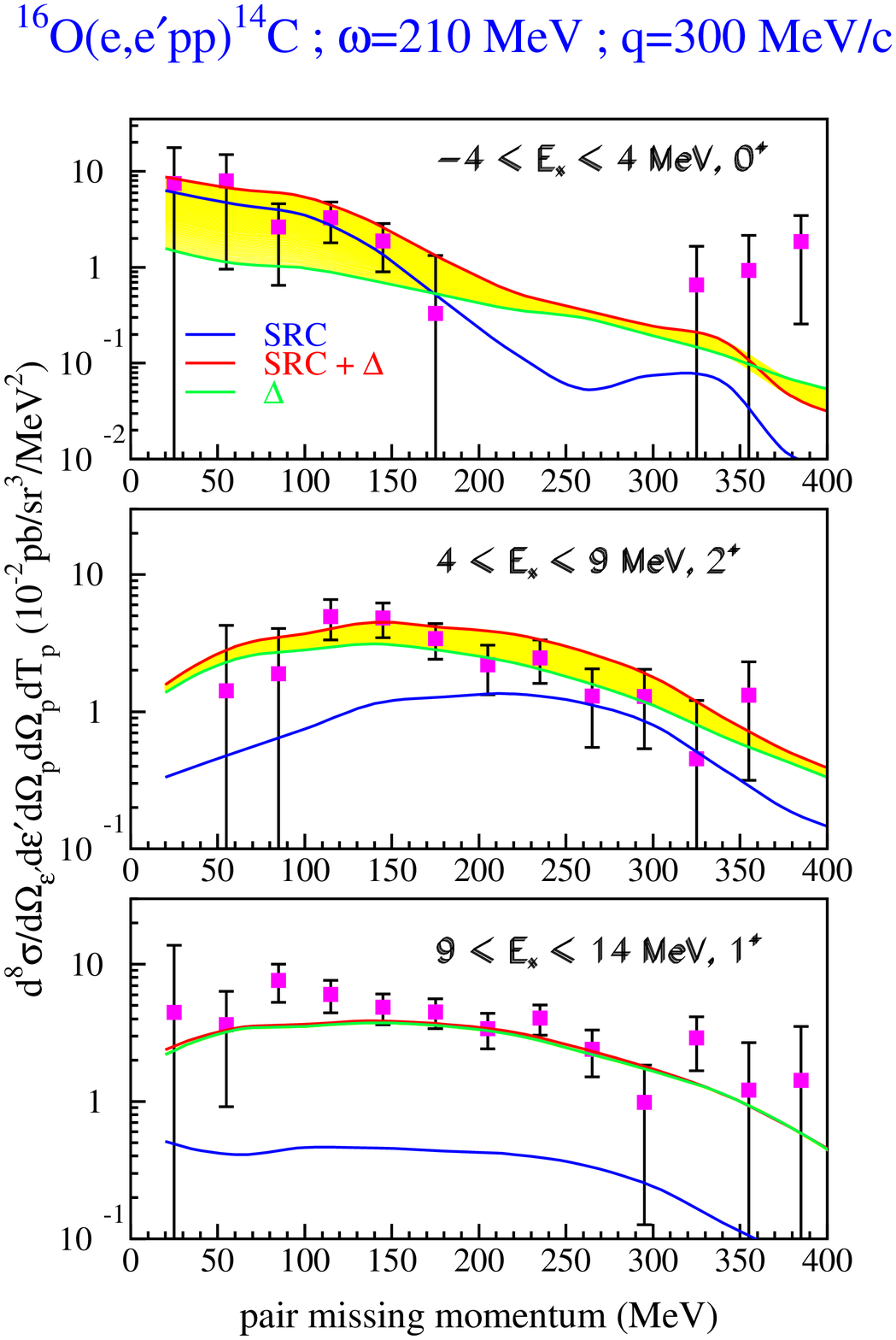}
\end{center}
\caption{\sl Calculated $^{16}$O($e,e'pp$) differential cross sections
for the kinematics of the NIKHEF experiment from Ref.~\protect
\cite{gercoprl}. The green curve is the calculated contribution from
intermediate $\Delta$ creation, ignoring all effects from SRC.  The
blue curve the contribution from the central correlations and the red
curve the coherent sum of the $\Delta$ and SRC contribution.}
\label{fig:nikhef}
\end{figure}

Figure \ref{fig:nikhef} shows the measured and calculated differential
cross sections versus missing momentum for the kinematics of the
$^{16}$O($e,e'pp$) NIKHEF experiment reported in Ref.~\cite{gercoprl}.
In this experiment the average energy and momentum transfer was
$\omega$=206~MeV and q $\approx$ 300 MeV/c.  A satisfactory agreement
between the calculations and the data is obtained using the
correlation function from Ref. \cite{gearhart}.  Remark that this
correlation function, denoted as ``G-matrix'' in Figure
\ref{fig:centralf}, could satisfactorily describe the
$^{12}$C$(e,e'pp$) data from Ref.~\cite{blom98} and the
$^{4}$He$(e,e'pp$) results contained in Fig.~\ref{fig:helium}.  The
central correlations are manifesting themselves in the $(e,e'pp)$
differential cross sections at low pair c.o.m. momenta.  Accordingly,
the central correlations are selecting diprotons for which the
c.o.m. angular momentum is determined by $\Lambda =0$ and the relative
motion by $^1S_0$ quantum numbers.  This selectivity makes the central
correlations to be an almost vanishingly small contribution for the
transition to the $1^+$ state that is dominated by relative $P$-wave
amplitudes and, consequently, intermediate $\Delta$ creation.
\begin{figure}[tbh]
\begin{center}
\includegraphics[bb=45 80 488 780,totalheight=10.cm,width=9.cm]{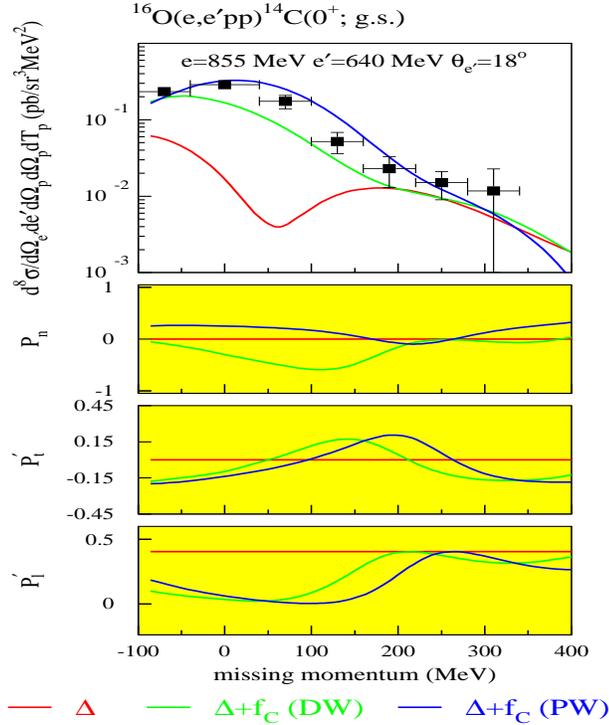}
\end{center}
\caption{\sl The missing momentum dependence of the
$^{16}$O($\vec{e},e' \vec{p}p$)$^{14}$C(0$^+$,g.s.) differential cross
section and recoil polarization observables in
superparallel kinematics.  The preliminary data are from
Ref.~\protect \cite{guenther99}. Both plane (PW) and distorted wave (DW)
calculations are shown.  The red curves include only $\Delta$
currents. The green and blue curves include $\Delta$ currents and  central
correlations.  The latter are implemented using the G-matrix correlation
function from Ref.~ \protect \cite{gearhart}.}
\label{fig:gun1}
\end{figure}
We now consider the reaction $^{16}$O($\vec{e},e' \vec{p}p$) in the
kinematics of a MAMI experiment that is presently in the process of
being analyzed \cite{guenther99,guenther}. These measurements have
resolved transitions to discrete states in the $^{14}$C residual
nucleus.  For this particular case we address the sensitivity of the
different observables to both the central and spin-dependent
correlations.  Furthermore, we want to demonstrate the power of using
recoil polarization measurements with and without polarized electrons.
The sensitivity to the final-state interactions is investigated in
Figure \ref{fig:gun1}.  In the asymmetric situation of super-parallel
kinematics, the distortions which the ejectiles undergo induce both a
shift and reduction of the plane-wave result for the differential
cross section.  The double polarization observables $P_t'$ and $P_l'$,
on the other hand, are only marginally affected by the FSI effects.
The results of Figure ~\ref{fig:gun1} include solely central
correlations.  The central correlations were implemented with the GD
correlation function that was also used in Figure ~\ref{fig:nikhef}.
Just as for the latter case, the effect of the central correlations is
restricted to the low pair missing (or c.o.m.)  momenta. The large
missing momentum tail is dominated by the $\Delta$ currents. Figure
\ref{fig:gun2} shows the results of a distorted wave calculation
including both central, spin-isospin and tensor correlations using the
correlation functions from the variational calculations shown in
Figure~\ref{fig:pieper}. The variational calculations produce a very
soft central correlation function and this results in a severe
underestimation of the $^{16}$O($e,e'pp$)$^{14}$C(0$^+$,g.s.)  data at
low pair missing momenta.  The inclusion of the state-dependent
correlations (spin-isospin and tensor) makes the calculated results to
move further away from the preliminary data as a destructive
interference between the state-dependent and the central correlations
is observed.  It appears that the variational calculations of
Ref~\cite{pieper} predict a central correlation functions that is
presumably too soft.  Inclusion of the state-dependent correlations,
which were consistently derived in the variational calculations, does
not seem to cure the observed deviation between the model predictions
and the preliminary data.  A similar qualitative deviation between the
predictions with the correlation functions from Ref.\cite{pieper} was
observed in the $^{12}$C$(e,e'pp)$ case (Ref.~\cite{blom98}).  The
power of the recoil polarization measurements in super-parallel
kinematics is illustrated in Figures \ref{fig:gun1} and
\ref{fig:gun2}.  The $P_n$ and $P_t'$ recoil polarization observable
are respectively determined by the $W_{LT}$ and $W'_{LT}$ structure
functions.  Accordingly, they vanish in the situation that only
(transverse) two-body currents contribute to the reaction process.
After inclusion of the correlations the recoil polarization
observables become relatively large.

\begin{figure}
\begin{center}
\includegraphics[bb=45 80 488 780,totalheight=10.cm,width=9.cm]{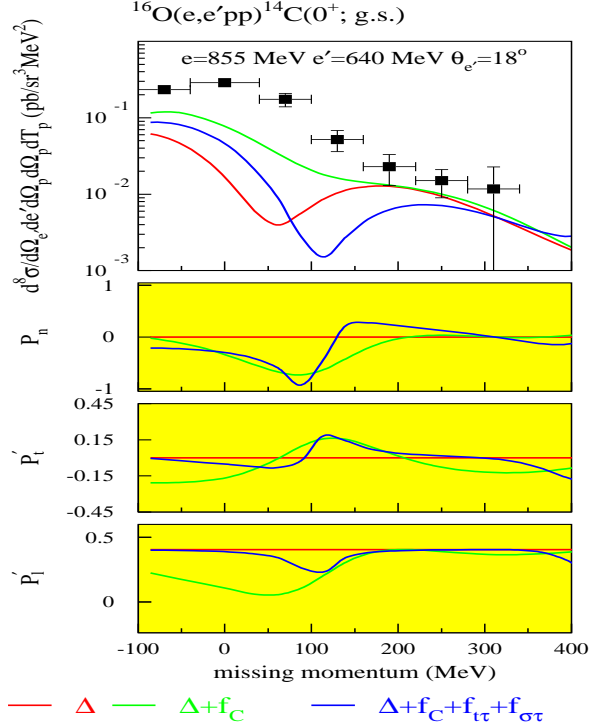}
\end{center}
\caption{\sl The sensitivity of the $^{16}$O($\vec{e},e'
        \vec{p}p$)$^{14}$C(0$^+$,g.s.) differential cross section and
        recoil polarization observables to the spin dependent
        correlation functions.}
\label{fig:gun2}
\end{figure}

\begin{figure}
\begin{center}
\includegraphics[bb=69 113 533 777,totalheight=9.cm,width=7.cm]{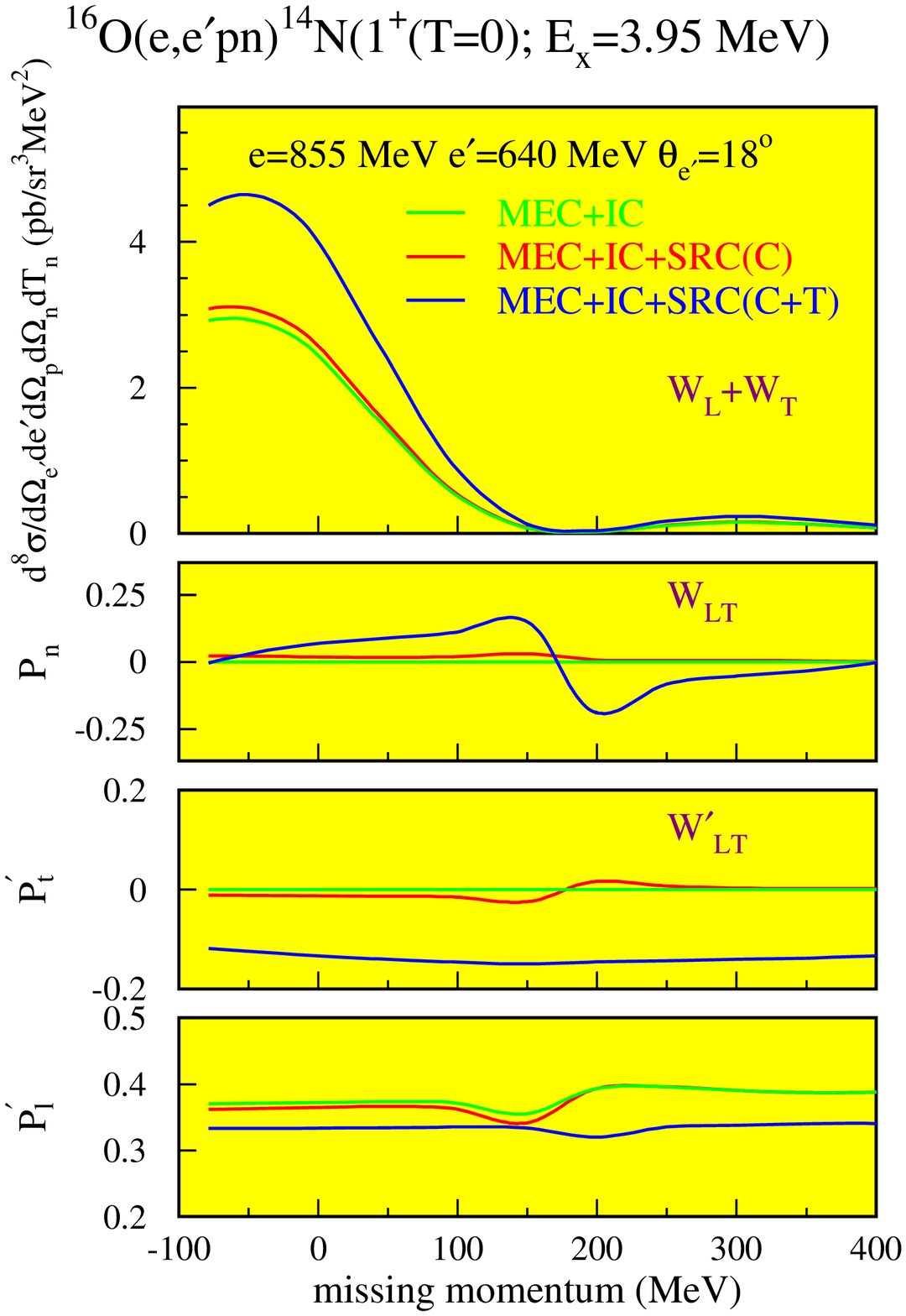}
\includegraphics[bb=69 113 533 777,totalheight=9.cm,width=7.cm]{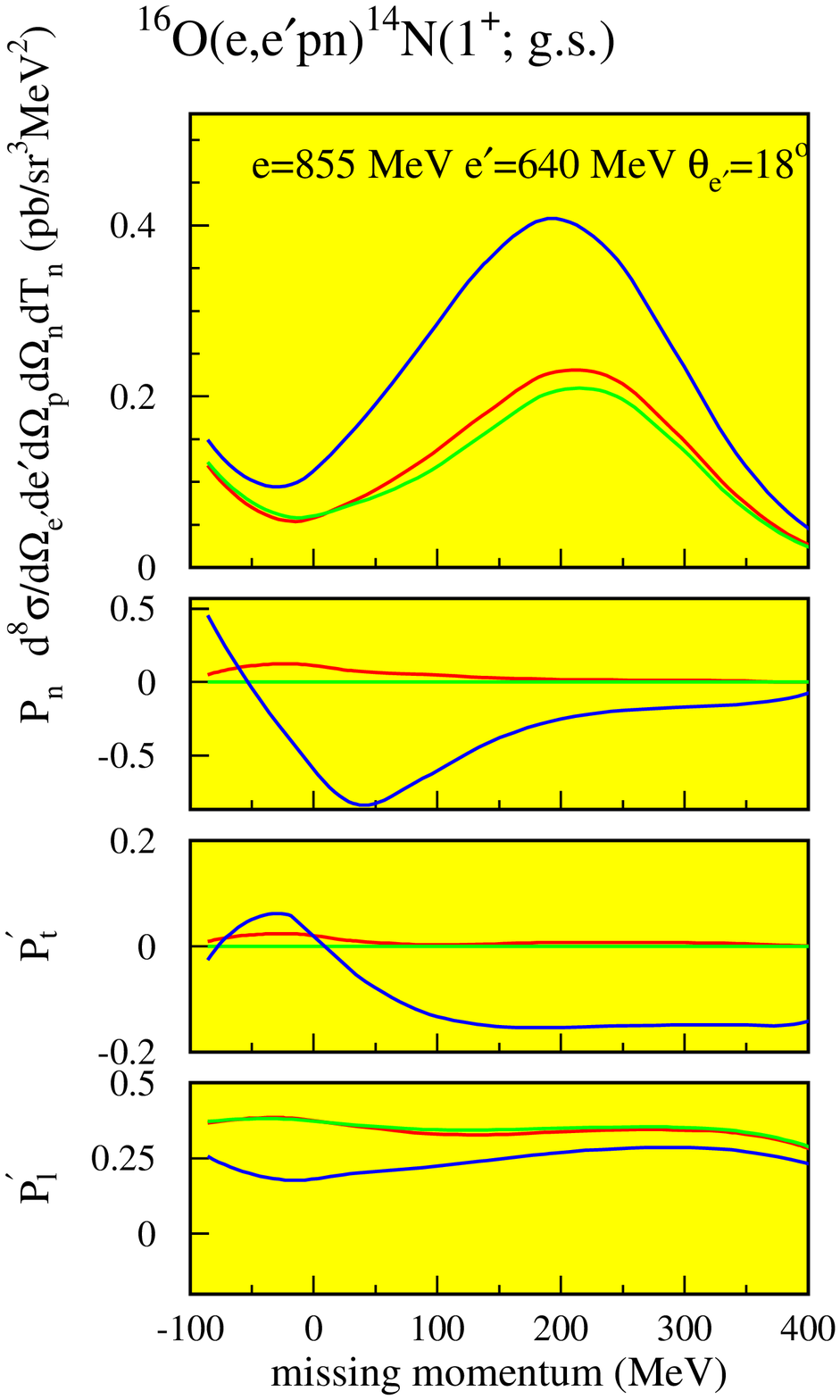}
\includegraphics[bb=69 113 533 777,totalheight=9.cm,width=7.cm]{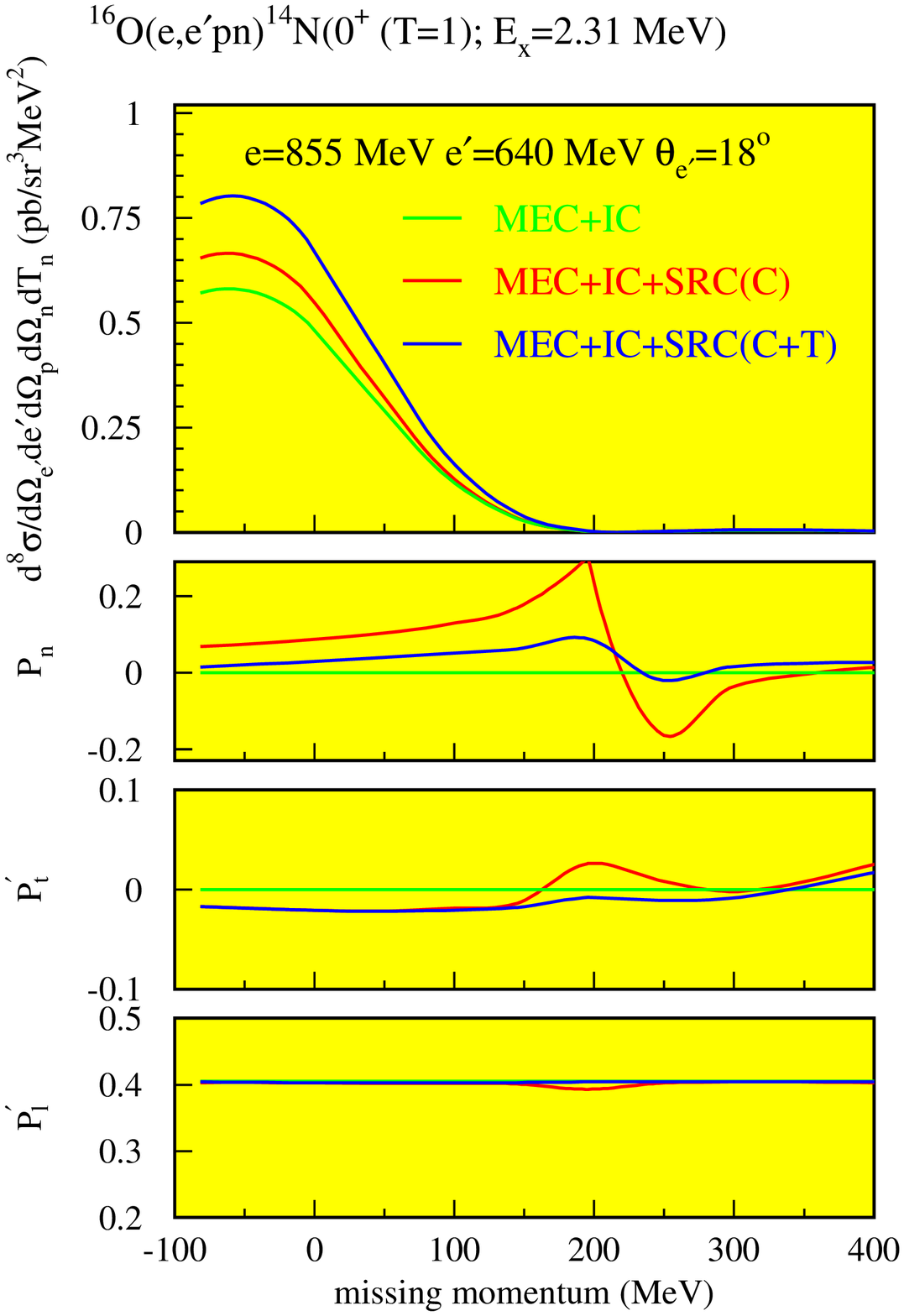}
\caption{\sl The missing momentum dependence of the $^{16}$O($e,e'pn$)
differential cross section and recoil polarization observables $P_n,
P'_l$ and $P'_t$ for excitation of the low lying states with two-hole
structure in $^{14}$N. The calculations refer to the situation in
which the proton (neutron) is detected parallel (anti-parallel) to the
direction of the momentum transfer.  The blue curve is calculated in
the distorted-wave approximation including the two-body currents,
central (``C'') and tensor (``T'') correlations. The green curve shows
the contribution from the two-body currents (``MEC+IC'').  The red
curves omit tensor correlations from the full calculation.}
\label{fig:eepn}
\end{center}
\end{figure}

Triple coincidence reactions of the type $A(e,e'pn$) are expected to
exhibit a major sensitivity to tensor correlations in nuclei, 
provided that one can separate the strength arising from one-body
photon coupling to a tensor correlated proton-neutron pair from
two-body current contributions, like meson-exhange contributions.  In
$(e,e'pp)$ studies at lower four-momentum transfer, a powerful tool in
the search for nucleon-nucleon correlations is the selectivity of the
final state with respect to the quantum numbers of the correlated
nucleon pair that actively participates in the reaction process
\cite{giusti98,ryckebusch99}.  To study whether a similar sort of
sensitivity exists for the tensor correlations, $^{16}$O($e,e'pn$)
results for excitation of specific states in $^{14}$N are presented.
Experiments of this type are scheduled at the Mainz electron
accelerator \cite{peterprop}.  Restricting ourselves to the (dominant) $p-shell$
components, the following selectivity for the quantum numbers of the
active proton-neutron pair emerges when considering $^{16}$O($e,e'pn$)
decay to the states with angular momentum $J_R$ in $^{14}$N
\begin{equation}
\begin{array}{ll}
J_R=0^+ & ^1S_0 (\Lambda=0), ^3P_1 (\Lambda=1)\\ J_R=1^+ & ^3S_1
(\Lambda=0,2), ^3P_{0,1,2} (\Lambda=1), \\ & ^1P_1 (\Lambda=1), ^3D_1
(\Lambda=0) \\ J_R=2^+ & ^1S_0 (\Lambda=2), ^3S_1 (\Lambda=0,2),
^3P_{1,2} (\Lambda=1), \\ & ^1D_2 (\Lambda=0), ^3D_2 (\Lambda=0)
\end{array}
\end{equation}
where the standard convention ($^{2S+1}l_J$) for the relative
two-nucleon wave function is adopted and $\Lambda$ is the orbital
quantum number corresponding with the c.o.m. motion of the pair.
Predictions for the exclusive $^{16}$O($e,e'pn$) cross sections are
shown in Figure~\ref{fig:eepn} for the two low-lying $1^+$ states
(respectively at 0. and 3.95~MeV excitation energy) and the $0^+$
state at $E_x=$2.31~MeV.  These states in the low-energy spectrum of
$^{14}$N are established to have a two-hole character relative to the
ground state of $^{16}$O \cite{snelgrove69} and are therefore expected
to be substantially populated in a direct $^{16}$O$(e,e'pn)$ reaction.
The two-hole overlap amplitudes employed in these calculations are
taken from Ref.~\cite{cohen70}.  The correlation functions are those
from \cite{pieper}.  We have selected so-called ``superparallel''
kinematics which makes the two nucleons to move along the momentum
transfer.  As is clearly reflected in the calculated cross sections of
Figure~\ref{fig:eepn}, the ground state has a mixed $\Lambda$=0,2
character, whereas the $J_R=1^+$ state at $E_x=$3.95~MeV has a clear
$\Lambda$=0 structure. As evidenced by the results of
Figure~\ref{fig:eepn} the effect of the tensor correlations in
exclusive proton-neutron knockout is sizeable.  The strongest
sensitivity to the tensor correlations is noticed in the peak of the
cross sections for the transition to the $1^+$ states.  In this
missing momentum domain the reaction is dominated by absorption on a
proton-neutron pair in a $^3S_1$ configuration. The effect of the
central correlations is small in comparison with the tensor
contributions.  For the transition to the $0^+(T=1)$ state, for which
no $^3S_1$ pair combination contributes, the effect of the central and
tensor correlations is modest and about equal. In this particular case
the proton-neutron knockout cross section is dominated by the two-body
currents.

Also shown in Fig.~\ref{fig:eepn} are the predictions for the double
recoil polarization observables ($P_l'$ and $P_t'$) that can be
determined in $(\vec{e},e'\vec{p}n)$ measurements. In superparallel
kinematics, $P_t'$ is uniquely determined by the $W_{LT}'$ structure
function.  The background of two-body current contributions to
proton-neutron knockout is exclusively transverse.  With no
ground-state correlations contributing, $P_t'$ would be extremely
small. The calculations learn that as soon as sizeable tensor
correlation effects come into play, the $P_t'$ becomes large.  In that
respect, the recoil polarization observable is a measure for the
tensor correlations which is relatively free of ambiguities with
respect to the final-state interaction.

\begin{figure}
\begin{center}
\includegraphics[bb=27 46 568 692,totalheight=12.cm,width=11.cm]{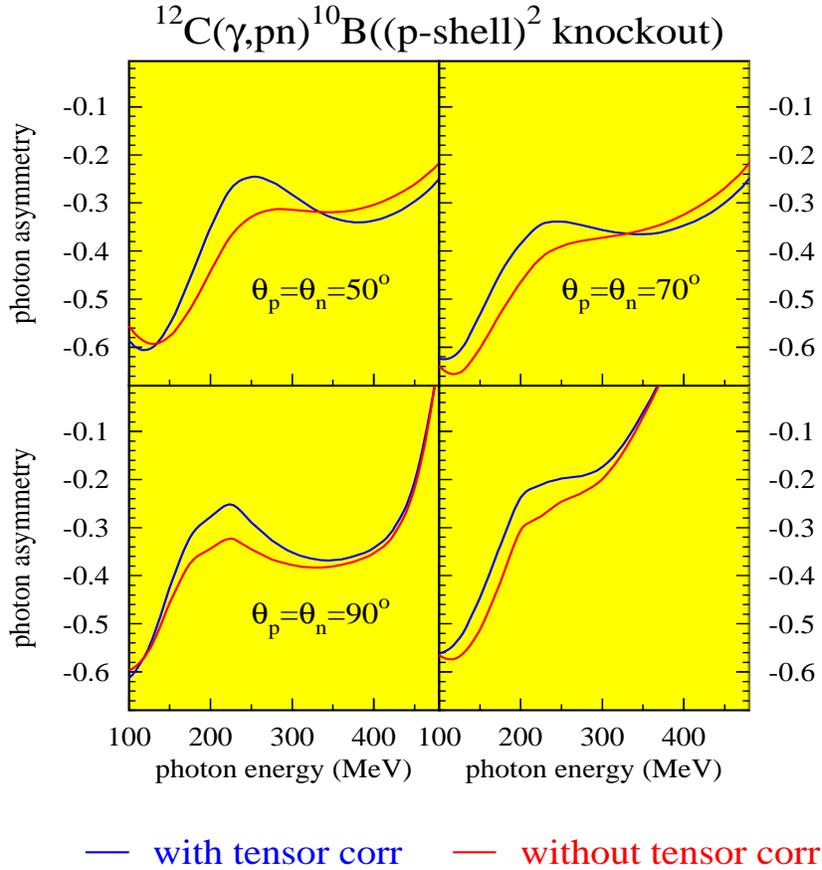}
\end{center}
\caption{\sl The photon asymmetry as a function of the photon-energy for
$(p-shell)^2$ knockout in $^{12}$C($\vec{\gamma},pn$) at various polar
angles in coplanar and symmetrical kinematics.  The red curve omits
tensor correlations from the full calculations that include the
meson-exchange currents, $\Delta$-isobar currents, central and
spin-dependent correlations.}
\label{fig:assym}
\end{figure}

Next, we investigate the role of tensor correlations in real photon
studies.  Figure~\ref{fig:assym} shows the the photon assymetry for
p-shell knockout in $^{12}$C$(\vec{\gamma},pn$) from 100 up to 440 MeV
photon energy. For these results coplanar and symmetrical kinematics
was selected.  A sizeable sensitivity to the tensor correlations is
observed.  We find the strongest effects in the photon energy range
200-300~MeV where the $\Delta$-isobar currents are heavily
contributing.  This behaviour points towards a strong interference
between tensor correlations and $\Delta$ currents.  A similar sort of
qualitative mechanism was observed by A. Fabrocini when studying the
role of the nucleon-nucleon correlations and two-body currents in inclusive $(e,e')$
\cite{fabrocini97}.

\begin{figure}
\begin{center}
\includegraphics[bb=69 11 470 605,totalheight=10.cm,width=9.cm]{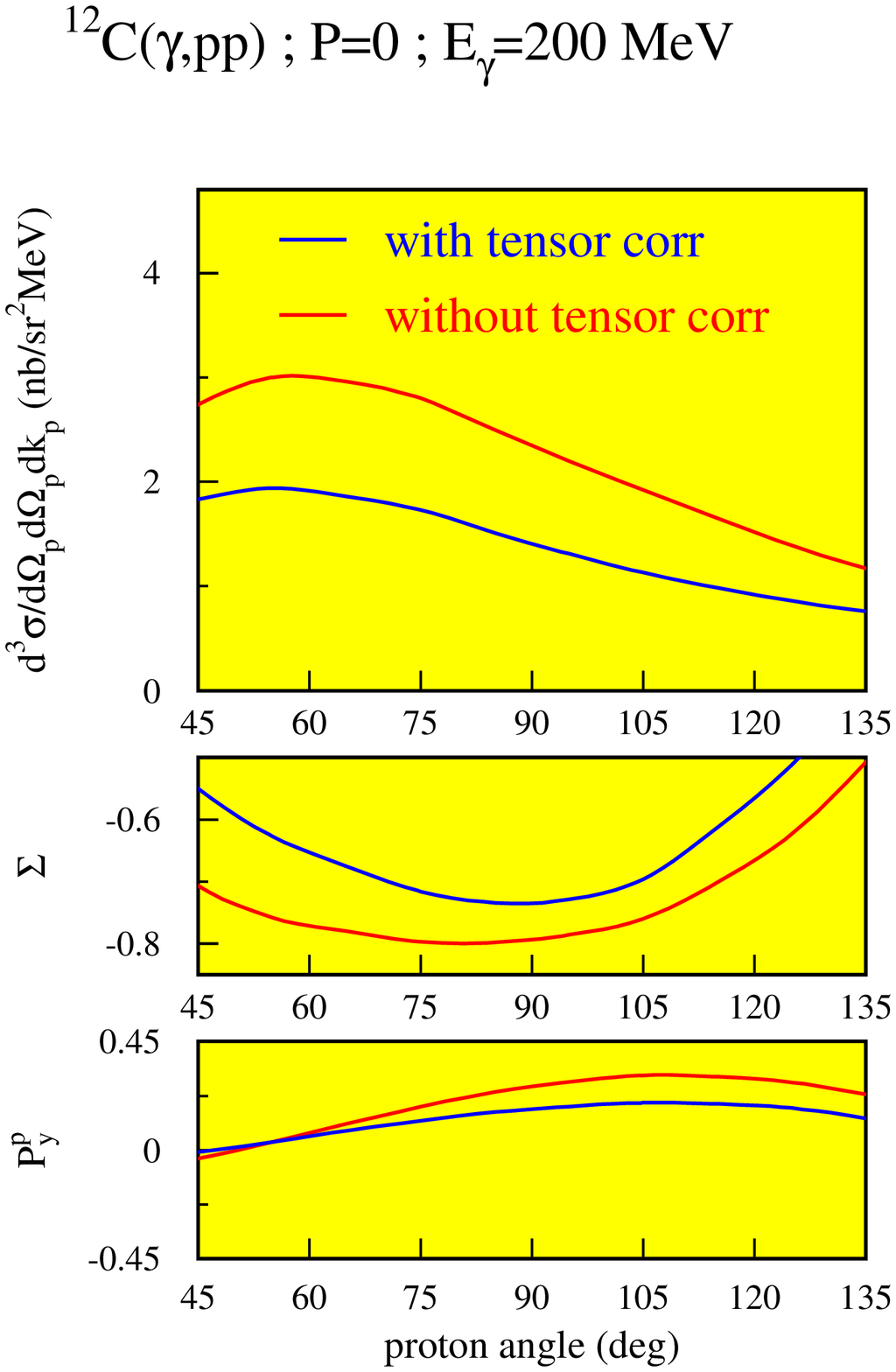}
\end{center}
\caption{\sl The differential cross section, photon asymmetry and analyzing
power for the reaction $^{12}$C($\gamma$,pp) as a function of the
polar angle for $E_{\gamma}$=200~MeV.  Quasi-deuteron kinematics is
selected. The red curve omits tensor correlations from the full
calculation that includes $\Delta$-currents, central and
spin-dependent correlations.} 
\label{fig:qdkinema}
\end{figure}

One could wonder whether this strong interference effect could be
further exploited to study tensor correlations in the medium.  The
$(\gamma,pp)$ channel was recently shown to be dominated by the
$\Delta$-isobar currents. Therefore, we address the question whether
tensor correlations could be probed in proton-proton knockout.  This
could a way of probing triplet $S$ diprotons, which is a unique medium
effect, and of learning more about the medium dependence of the tensor
correlations.  As displayed in Figure~\ref{fig:qdkinema} the effects
of the tensor correlations in the $(\gamma,pp)$ channel are
significant. In the differential cross section the tensor correlations
induce a substantial global reduction which exhibits a weak proton angle
dependence.  Also for the photon asymmetry the tensor correlations are
important.  As the photon asymmetry is far less subject to
uncertainties with respect to the FSI than the cross sections, it
represents a promising method for studying tensor correlations in the
medium.

\section{Summary}
A framework in which the effect of central, tensor and spin-isospin
NN-correlations on $A(e,e'NN)$ and $A(\gamma,NN)$ cross sections and
polarization observables can be calculated is available.  In this
formalism,  also the ``background'' two-nucleon knockout strength
attributed to meson-exchange and $\Delta_{33}$ currents is incorporated.
The results indicate that short-range central correlations produce
sizeable contributions to $A(e,e'pp)$.  The major effects are observed
at low missing ``diproton'' momenta.  This observation supports the
picture that the central short-range correlations result in hard
back-to-back collisions producing relatively high relative momenta
$p_{rel} >>$ and small c.o.m. momenta $P \approx 0$ in the correlated
part of the spectral function $P(\vec{k},E)$.  The tensor correlations, on the
other hand, appear in a wider range of pair c.o.m. momenta, which
reflects the fact that they are of intermediate range.  The results of
the calculations indicate that $A(\vec{e},e'\vec{p}N$) polarization
observables offer the possibility of effectively isolating the
longitudinal channel, creating favorable conditions to study the
central and spin-dependent correlations with ``minimized'' FSI
effects.  For these studies, superparallel kinematics (both nucleons
are ejected along the momentum transfer) reduces the systematic
uncertainties as it is very selective with respect to the structure
functions that are contributing to each of the observables.  The
results indicate that tensor correlations are {\em STRONGLY}
contributing to the proton-neutron knockout channel $A(e,e'pn)$.  Here,
the sensitivity to central short-range correlations is rather weak.
The $A(e,e'pn$) channel carries stronger signals
from the nucleon-nucleon correlations than $A(e,e'pp$) in comparable
kinematics.  A strong interference between the $\Delta$ current and
tensor correlations is found.  This feature makes the $A(\vec{\gamma},NN)$
asymmetries a promising variable to study tensor correlations in the
medium.

\noindent{\bf Acknowledgments}\\[1ex]
This work has been supported by grants from the Fund for Scientific
Research - Flanders (FWO) under grant number CRG970268. We want to thank
the organizers of this workshop for the wonderful days of physics in
Granada!

\end{document}